\documentclass[a4paper,fleqn,usenatbib]{mnras}

\usepackage{newtxtext,newtxmath}
\usepackage[T1]{fontenc}
\usepackage{ae,aecompl}

\usepackage{graphicx}	% Including figure files
\usepackage{amsmath}	% Advanced maths commands
\usepackage{amssymb}	% Extra maths symbols

\title[Molecular oxygen abundance]{Influence of galactic arm scale dynamics on the molecular composition of the cold and dense ISM II. Molecular oxygen abundance}

\author[Wakelam et al.]{
V. Wakelam$^{1}$, M. Ruaud$^{2}$, P. Gratier$^{1}$, I. Bonnell$^3$ 
\thanks{E-mail: valentine.wakelam@u-bordeaux.fr}
\\
$^{1}$Laboratoire d'astrophysique de Bordeaux, Univ. Bordeaux, CNRS, B18N, all\'ee Geoffroy Saint-Hilaire, 33615 Pessac, France\\
$^2$ NASA Ames Research Center, Moffett Field, CA, USA\\
$^3$ Scottish Universities Physics Alliance (SUPA), School of Physics and Astronomy, University of St. Andrews, North Haugh,
St Andrews, Fife KY16 9SS, UK
}

\date{Accepted XXX. Received YYY; in original form ZZZ}

\pubyear{2019}

\begin{document}
%\label{firstpage}
\maketitle
\begin{abstract}
Molecular oxygen has been the subject of many observational searches as chemical models predicted it to be a reservoir of oxygen. 
Although it has been detected in two regions of the interstellar medium, its rarity is a challenge for astrochemical models. In this paper, we have combined the physical conditions computed with smoothed particle hydrodynamics (SPH) simulations with our full gas-grain chemical model Nautilus, to study the predicted O$_2$ abundance in interstellar material forming cold cores. We thus follow the chemical evolution of gas and ices in parcels of material from the diffuse interstellar conditions to the cold dense cores. Most of our predicted O$_2$ abundances are below $10^{-8}$ (with respect to the total proton density) and the predicted column densities in simulated cold cores is at maximum a few $10^{14}$~cm$^{-2}$, in agreement with the non detection limits. This low O$_2$ abundance can be explained by the fact that, in a large fraction of the interstellar material, the atomic oxygen is depleted onto the grain surface (and hydrogenated to form H$_2$O) before O$_2$ can be formed in the gas-phase and protected from UV photo-dissociations. We could achieve this result only because we took into account the full history of the evolution of the physical conditions from the diffuse medium to the cold cores. 
\end{abstract}


%%%%%%%%%%%%%%%%%%%%%%%%%%%%%%%%%%%%%%%%%%%%%%%%%%

%%%%%%%%%%%%%%%%% BODY OF PAPER %%%%%%%%%%%%%%%%%%

\section{Introduction}

Oxygen and carbon are the two most abundant elements (after hydrogen and helium) in the universe 
and they are basic material for the organic matter. Astrochemical models predict that, in the cold
 molecular gas of the interstellar medium, carbon monoxyde (CO) constitutes the reservoir of volatile carbon 
 while water (frozen on interstellar dust) is the reservoir of oxygen. These two species are indeed 
 observed to contain large fractions of elemental oxygen and carbon. The third reservoir of oxygen is 
 predicted to be molecular oxygen (O$_2$). Pure gas-phase models predict an abundance of a few $10^{-5}$ 
 \citep[using the kida.uva.2014 public network,][]{2015ApJS..217...20W} under typical cold core conditions. 
 Such models even predict O$_2$ to be more abundant than H$_2$O as they do not include any formation of 
 water on interstellar dust (a very efficient process). Chemical models including grain surface processes 
 (and gas-grain interactions) predict gas-phase abundances of O$_2$ much smaller as part of the oxygen 
 can be depleted onto the grain surfaces and be hydrogenated (forming H$_2$O). The peak abundance of 
 gaseous O$_2$ however usually reaches $\sim 10^{-6}$ for a cloud age between $10^5$ and $10^6$~yr 
 \citep[see for instance][]{2011A&A...530A..61H}. \\
 Based on these model predictions, molecular oxygen has been intensively searched in the interstellar medium, 
 first with ground based observations of the isotopic form $^{16}$O$^{18}$O 
 \citep[see][and references therein]{Pagani2003} and then with the SWAS satellite 
 \citep{2000ApJ...539L.123G}. All these studies reported a non detection of the molecule and 
 upper limits on the abundance smaller and smaller with time. Later, with the Odin and then Herschel 
 space telescopes, detections have been reported but they have shown the very small abundance of 
 O$_2$ and a scarcity of detections. Molecular oxygen has been detected only in two sources: Orion 
 \citep{2011ApJ...737...96G} and $\rho$ Oph A \citep{2007...466..999L,2012...541A..73L} 
 with an abundance between two and three orders of magnitude smaller than what is predicted by astrochemical models.  Since these detections, O$_2$
  has been searched for in many objects: CO-depleted cold cores \citep{2016ApJ...830..102W}, 
  Sgr A foreground absorption \citep{Sandqvist2015}, low mass protostars 
  \citep{2013...558A..58Y,2018...618A..11T}, the Orion bar \citep{2012ApJ...752...26M} 
  without success. More over, additional observations of the Orion region seem to indicate that the O$_2$ detection there 
  \citep{2011ApJ...737...96G} is associated to a shock region rather than a cold core \citep{2014ApJ...793..111C}. 
  Interestingly, the other region where O$_2$ has been detected, $\rho$ Oph A, is also the only region 
  where both O$_2$H and HOOH have been detected \citep{2011A&A...531L...8B,2012A&A...541L..11P}. 
  A list of observed column densities and abundances as well as upper limits are given in Table~\ref{table_obs}. This list is not meant to be exhaustive but gives an overview of the most recent observations in a variety of sources. \\
  Many theoretical studies have tried to explain this result exploring a decrease of the 
  elemental abundance of oxygen \citep{2011A&A...530A..61H}, the presence of bistabilities in the model 
  results \citep{2001A&A...370..557V}, and the impact of the value of the rate coefficient for the main 
  reaction forming O$_2$ \citep[O + OH $\rightarrow$ O$_2$ + H,][]{1997A&A...324..221M,2008ApJ...681.1318Q}. \\
 All chemical models presented in the literature and focusing on O$_2$ formation assume very simple 
 physical models (mostly static). In this paper, we will show the impact of the cloud formation history on the 
 predicted O$_2$ abundances. The model presented here have already been used in \citet{2018A&A...611A..96R} 
 to show that considering the full history of the physical conditions during the formation of cold cores 
 from the diffuse medium has a strong impact on the cloud composition changing the gas-phase C/O global elemental ratio and the 
 electron fraction.

 \begin{table*}
\caption{Column densities and/or abundances of O$_2$ (detection and upper limits) from the literature.}
\begin{center}
\begin{tabular}{llll}
\hline
\hline
Source & Column density (cm$^{-3}$) & Abundance (with respect to H$_2$) & Reference \\
\hline
TMC1 - NH$_3$ & $<6.8\times 10^{14}$ & $< 7.7\times 10^{-8}$ & \citet{Pagani2003} \\ 
L134N - NH$_3$ &  $<1.1\times 10^{15}$ & $< 1.7\times 10^{-7}$ & \citet{Pagani2003} \\ 
L429 & $< 1.1\times 10^{16}$ & $< 9.2\times 10^{-8}$ & \citet{2016ApJ...830..102W} \\
Oph D & $< 1.2\times 10^{16}$ & $< 1.1\times 10^{-7}$ & \citet{2016ApJ...830..102W} \\
L1544 & $< 8.2\times 10^{15}$ & $< 6.3\times 10^{-8}$ & \citet{2016ApJ...830..102W} \\
L694-2 & $< 1.6\times 10^{16}$ & $< 1.6\times 10^{-7}$ & \citet{2016ApJ...830..102W} \\
Sgr A & $<1.4\times 10^{16}$ & $< 5\times 10^{-8}$ & \citet{Sandqvist2015} \\
IRAS4 (protostar) &  $< 1.2\times 10^{15}$ & $< 5.7\times 10^{-9}$ & \citet{2013...558A..58Y} \\
IRAS4 (cloud) &  $ (2.8 - 4.3)\times 10^{15}$* & $(4.3-2.2)\times 10^{-7}$* & \citet{2013...558A..58Y} \\
Orion bar & $< 1\times 10^{16}$ &  - & \citet{2012ApJ...752...26M} \\
IRAS16293-2422 & $<1.7\times 10^{15}$ & $< 1.2\times 10^{-7}$ & \citet{Pagani2003} \\ 
IRAS16293-2422 & $< 9\times 10^{19}$ &  - & \citet{2018...618A..11T} \\
$\rho$ Oph A & $<3.4\times 10^{15}$ & $<9.3\times 10^{-8}$ & \citet{Pagani2003} \\ 
$\rho$ Oph A & $10^{15}$ & $5\times 10^{-8}$ & \citet{2007...466..999L}\\
$\rho$ Oph A & $(3-6)\times 10^{15}$ & $\sim 5\times 10^{-8}$ & \citet{2012...541A..73L}\\
Orion & $6.5\times 10^{15}$ & $(0.3-7.3)\times 10^{-6}$ & \citet{2011ApJ...737...96G}\\
Orion A &  $<1.9\times 10^{15}$ & $< 8.9\times 10^{-8}$ & \citet{Pagani2003} \\ 
NGC2071 & $<2.6\times 10^{15}$ & $<1.5\times 10^{-7}$ & \citet{Pagani2003} \\ 
NGC6334I &  $<5.0\times 10^{15}$ & $<7.1\times 10^{-8}$ & \citet{Pagani2003} \\ 
(G0.26 - 0.01) & $<5.6\times 10^{16}$ & $<7.6\times 10^{-7}$ & \citet{Pagani2003} \\ 
M17SW & $<7.3\times 10^{15}$ & $< 5.7\times 10^{-7}$ & \citet{Pagani2003} \\ 
S68FIRS1 &  $<1.6\times 10^{15}$ & $< 9.7\times 10^{-8}$ & \citet{Pagani2003} \\ 
G34.3 + 0.2 &  $<5.2\times 10^{15}$ & $< 5.2\times 10^{-8}$ & \citet{Pagani2003} \\ 
\hline
* claimed to be a tentative detection by the authors.
\end{tabular}
\end{center}
\label{table_obs}
\end{table*}%

\section{Models description}

To simulate the chemistry during the formation of cold cores, we used the same methodology as \citet{2018A&A...611A..96R} with an up-to-date version of the chemical model Nautilus \citep{2016MNRAS.459.3756R}. The chemical evolution was computed for parcels of material evolving in a Galatic arm, which physical conditions (gas temperature, density, and visual extinction) were computed with the 3D SPH model from \citet{2013MNRAS.430.1790B}. The parcels of material were selected as follows. At the end of the simulations, maximum density peaks were identified (with a density larger than $10^5$~cm$^{-3}$). Then, SPH particles in a sphere of 0.5 pc in radius around this peak were identified and the past and future history of the physical conditions of these identified particles were extracted. The number of particle per cloud is not constant but ranges between 150 and 350. In total, we have about 2918 trajectories for nine clouds. These time dependent physical conditions were then used as inputs to the Nautilus gas-grain model. There is no feedback of the chemistry computed with Nautilus on the physical model. The dust temperature is computed following \citet{1991ApJ...377..192H}. More details on the method can be found in \citet{2018A&A...611A..96R}. As the physical model does not include self-gravity, the clouds do not undergo collapse but break after the peak density. For the present study, we have used the data up to the density peak and ignored the dispersion of the cloud after it. \\
The Nautilus gas-grain model is used in its 3-phase version, meaning that the chemistry is computed in the gas-phase and on the surface of the interstellar grains but making a distinction between the most external 2 monolayers of molecules (called surface) and the rest on the ice below (called bulk). Species from the gas-phase are physisorbed on the surfaces upon collision with grains. In addition to the surface and bulk reactions, the model computes desorption: thermal desorption, cosmic-ray induced desorption \citep[using the formalism of][]{1993MNRAS.261...83H}, chemical desorption \citep[using the formalism of][]{2007A&A...467.1103G}, and photodesorption with a yield of $10^{-4}$ for all desorptions \citep[see][for discussions]{2016MNRAS.459.3756R}. All the details of the equations and values of surface parameters can be found in a dedicated paper on Nauitlus: \citet{2016MNRAS.459.3756R}. \\
With respect to \citet{2018A&A...611A..96R}, the chemical networks have been updated based on the reactions proposed in  \citet{2017MNRAS.470.4075L}  and the binding energies proposed in \citet{2017MolAs...6...22W}. The initial abundances are the same as in \citet{2018A&A...611A..96R} (see their Table 1). 

\section{Predicted O$_2$ gas-phase abundance}
\begin{figure}
\includegraphics[width=1\linewidth]{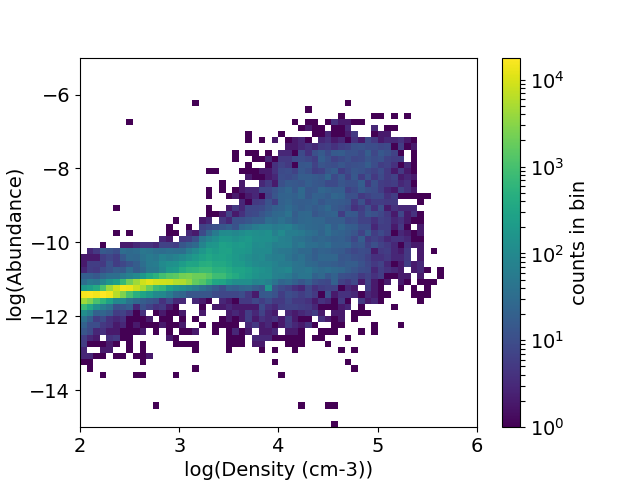}
\includegraphics[width=1\linewidth]{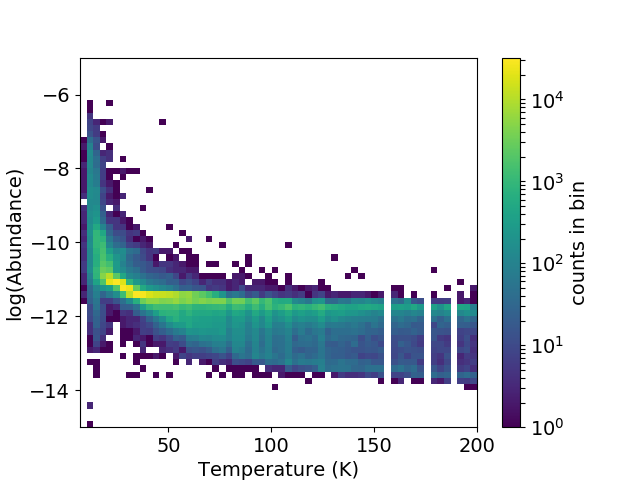}
\caption{2D histograms of the O$_2$ gas-phase abundance as a function of the density (up) and gas temperature (low) for our 2918 trajectories. \label{O2_param}}
\end{figure}

\begin{figure}
\begin{center}
\includegraphics[width=1\linewidth]{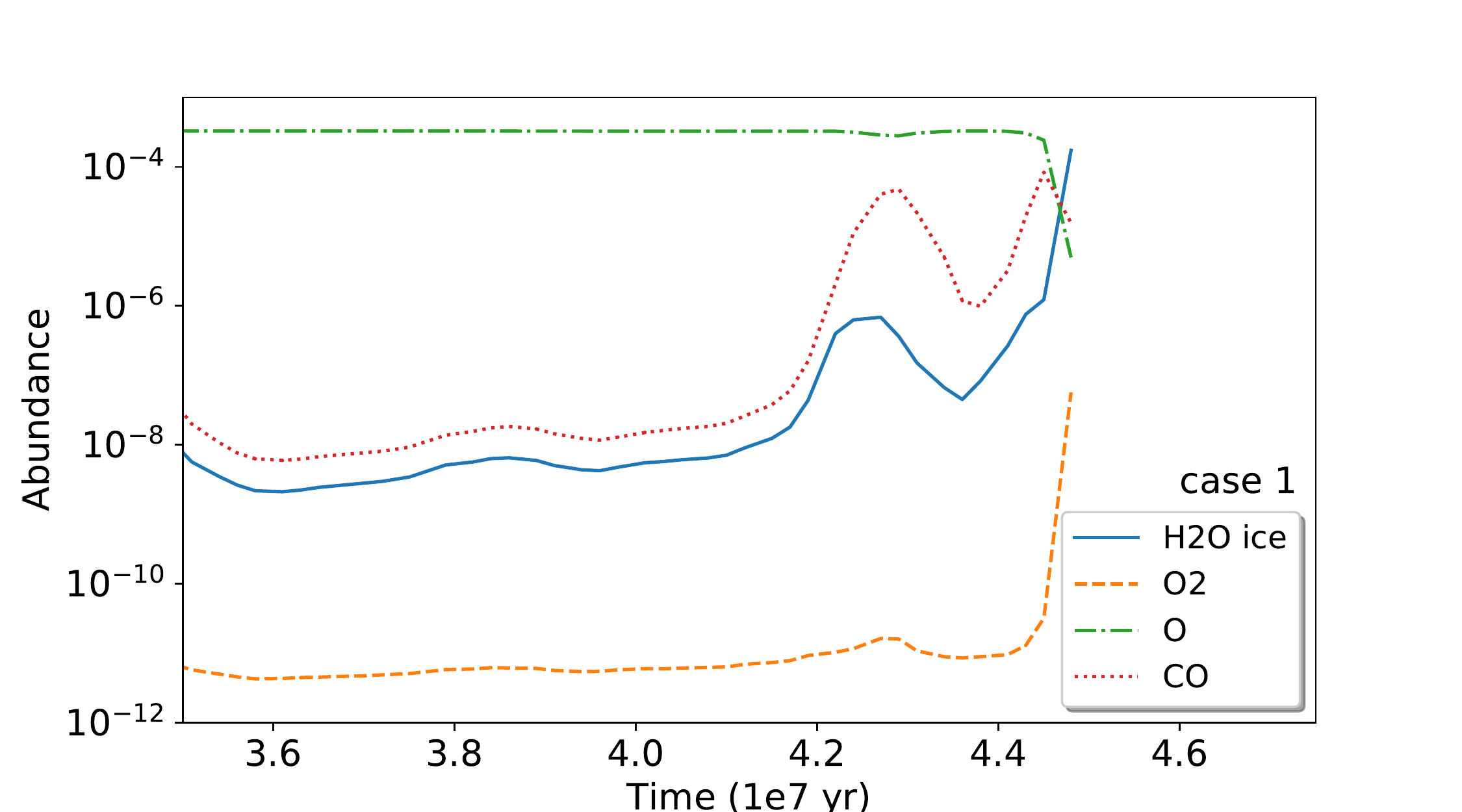}
\includegraphics[width=1\linewidth]{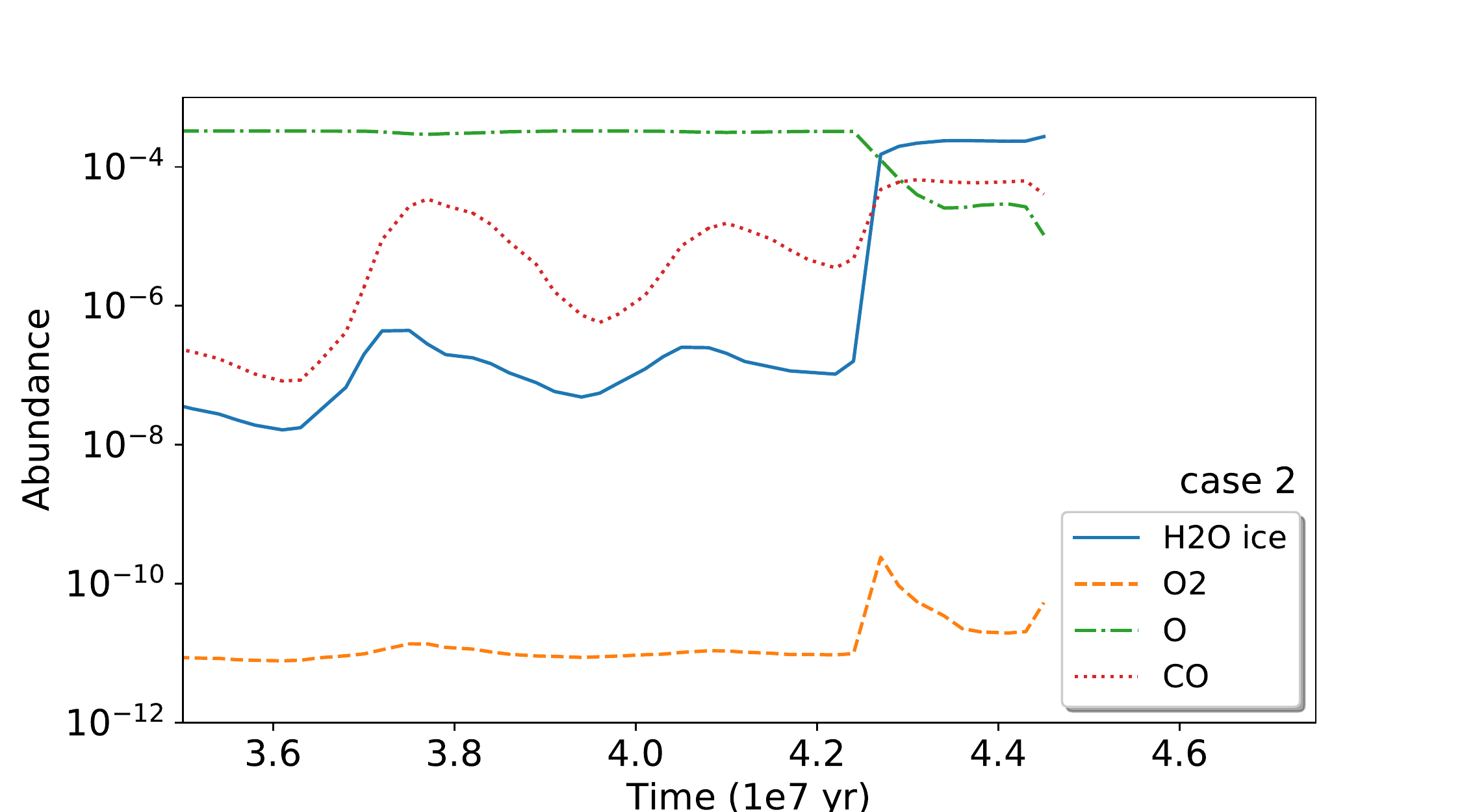}
\includegraphics[width=1\linewidth]{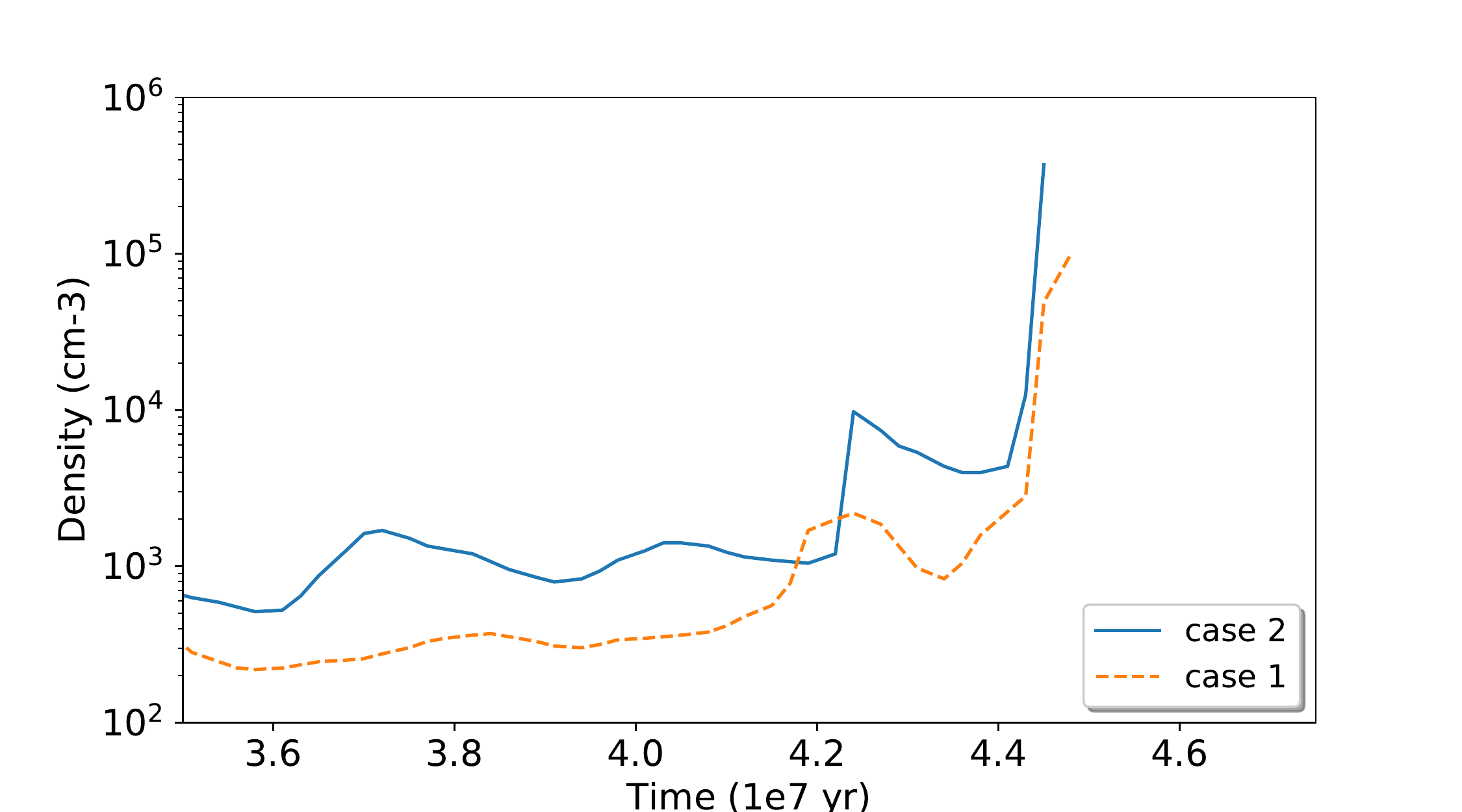}
\includegraphics[width=1\linewidth]{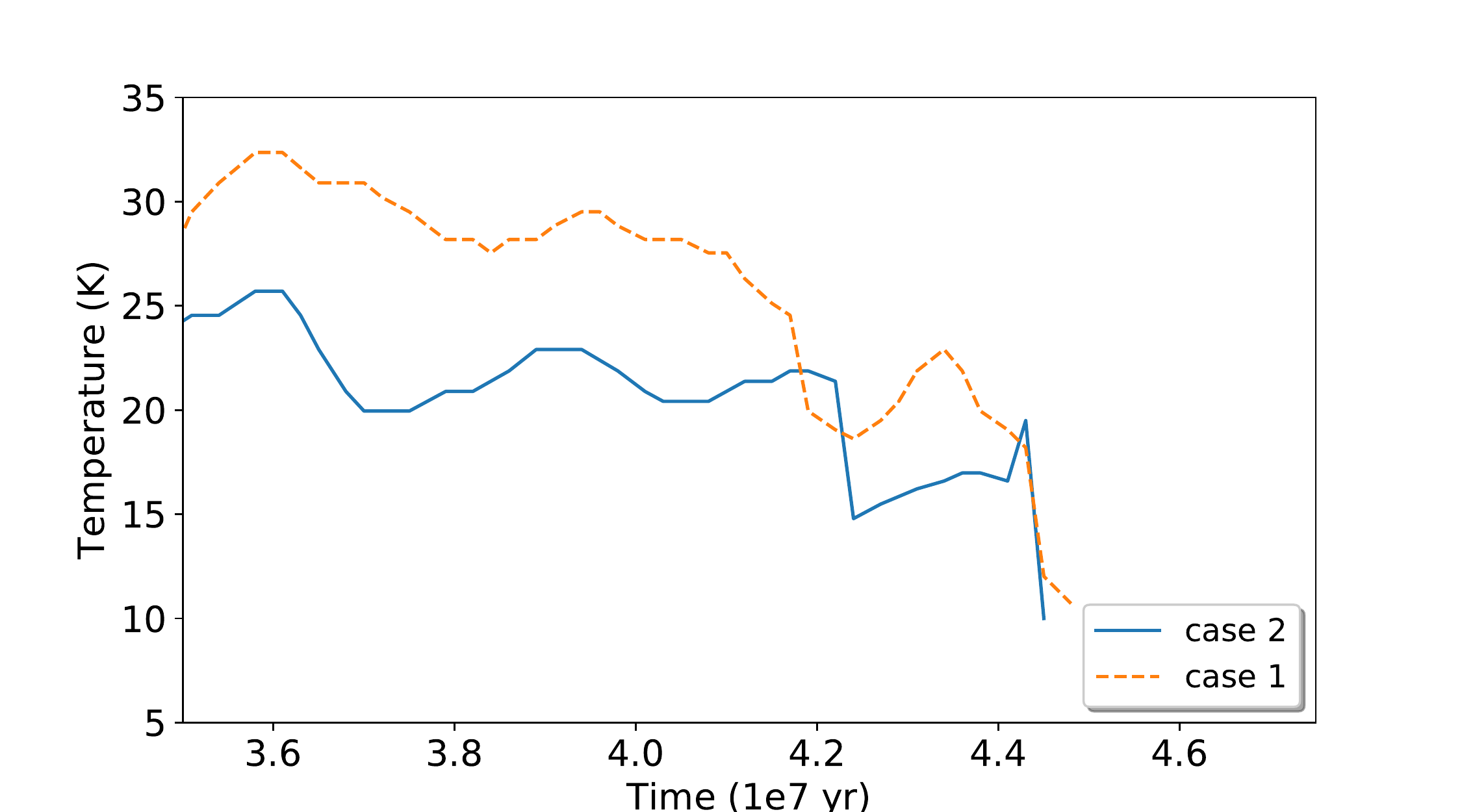}
\caption{Abundances of H$_2$O ice, gas-phase O$_2$, O, and CO, temperature and density as a function of time for cases 1 and 2.  \label{ab_clouds}}
\end{center}
\end{figure}

In Fig.~\ref{O2_param}, we plot the O$_2$ gas-phase abundance (with respect to H) predicted by our model as a function of 
the density and the temperature for all selected trajectories, whatever the time of the simulation (up to the density peak). 
The simulations include smaller densities and higher temperatures but we have zoomed to the highest densities 
where molecules form. For the same physical conditions, the O$_2$ 
gas-phase abundances spread over a large range of values covering several orders of magnitude. The maximum abundance is below $10^{-6}$ with the majority of the points laying below $10^{-8}$. \\
To understand this result, we have selected two trajectories producing either a "large" peak abundance (of a few $10^{-8}$ - case 1) 
or a small peak abundance (of a few $10^{-11}$ - case 2) at the maximum peak density (of $\sim 10^5$~cm$^{-3}$ in both cases).  
Fig.~\ref{ab_clouds} shows the time dependent abundances of gas-phase O$_2$, O and CO, and H$_2$O ice for these two trajectories. In all our cases, molecular oxygen is formed in the gas-phase by the neutral-neutral reaction O + OH $\rightarrow$ O$_2$ + H while OH is formed by the dissociative recombination of protonated water. Over the majority of the simulation, the low density allows for the photo-dissociation of O$_2$ by UV photons. Fig.~\ref{ab_clouds} also shows the gas temperature and density as a function of time for these two trajectories. In case 2, the material undergo a first phase of large density of about $10^4$~cm$^{-3}$ approximately $2\times 10^6$~yr before the density peak. During this transition, a large fraction of the oxygen is depleted onto the grains and hydrogenated to form H$_2$O. The abundance of H$_2$O ice in case 2 is about $3\times 10^{-4}$ at $4.3\times 10^7$~yr of the simulation while it reaches this value in case 1 only at the peak density (at $4.5\times 10^7$~yr).  As a consequence, the fraction of oxygen available for the gas-phase chemistry at the density peak is much smaller than in case 1. In case 1, the O$_2$ ice abundance is about $10^{-8}$ while it is negligible in case 2 at the density peak. The overall small abundance of gas-phase O$_2$ in Fig.~\ref{O2_param} can then be explained by the fact that the depletion of the oxygen (followed by hydrogenation) is faster in many trajectories (because of the time scale of evolution of the physical conditions) than the formation of O$_2$ in the gas-phase. In the case with a larger O$_2$ gas-phase abundance, the density conditions do not allow for an efficient oxygen depletion before the density peak. \\

\begin{figure*}
\includegraphics[width=0.33\linewidth]{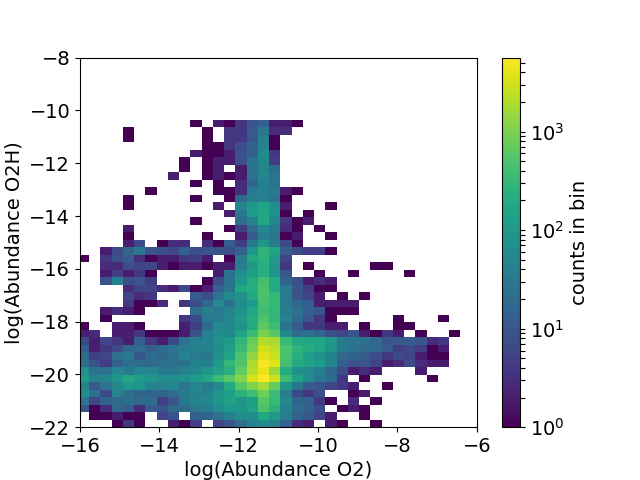}
\includegraphics[width=0.33\linewidth]{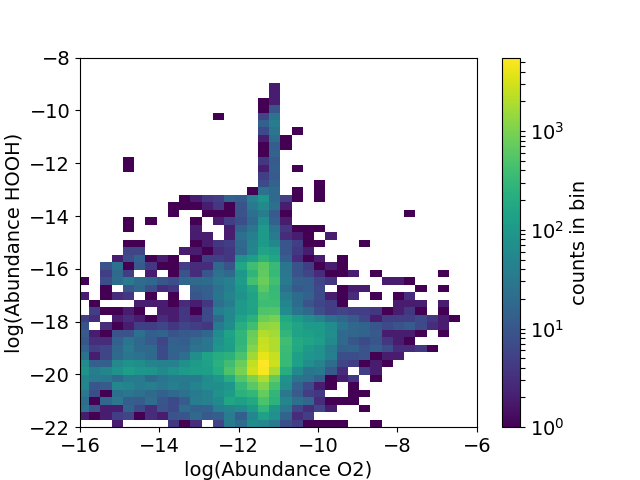}
\includegraphics[width=0.33\linewidth]{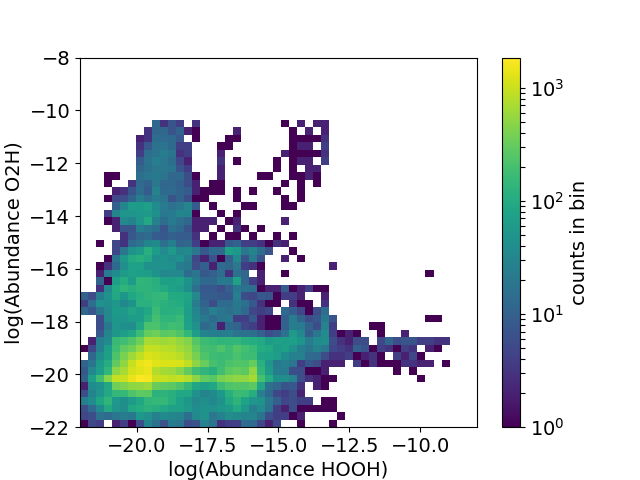}
\caption{2D histograms of gas-phase abundances of HOOH, O$_2$H and O$_2$. \label{O2_O2H_H2O2}}
\end{figure*}

\begin{figure}
\includegraphics[width=1\linewidth]{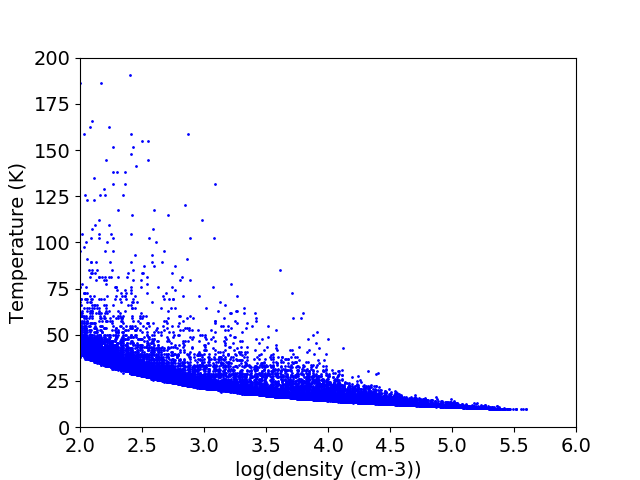}
\caption{Total proton density (in cm$^{-3}$) as a function of temperature used in the simulations. \label{points_dens_temp}}
\end{figure}

Considering the fact that, one of the only two regions where O$_2$ has been detected is also the only region where HOOH and O$_2$H have been detected, we have looked at the model prediction for these two species. Fig.~\ref{O2_O2H_H2O2} shows the 2D histograms of these species abundances (O$_2$ versus O$_2$H, O$_2$ versus HOOH, and HOOH versus O$_2$H). In these simulations, they do not seem to be correlated. Both O$_2$H and HOOH are predicted by the model to form on the grains by hydrogenation of O$_2$. Molecular oxygen first has to be formed in the gas-phase and accrete on the grains. Then, successive hydrogenation of O$_2$ produces partial evaporation of first O$_2$H and then HOOH through chemical desorption. This time dependent sequence can explain why there are some points associated with larger O$_2$H and lower O$_2$ (O$_2$ has to deplete for O$_2$H to form) and some points with large HOOH and lower O$_2$H (O$_2$H is transformed into HOOH). 
From a general point of view, O$_2$ gas-phase abundance larger than $10^{-8}$ are associated with negligible abundances of O$_2$H and HOOH. In $\rho$ Oph, the observed abundances of O$_2$ is estimated to be $\sim 5\times 10^{-8}$ (with respect to H$_2$, see Table~\ref{table_obs}) and the HOOH and O$_2$H abundances to be $\sim 10^{-10}$ \citep{2011A&A...531L...8B,2012A&A...541L..11P}.
 The observed HOOH and O$_2$H cannot be reproduced by our models contrary to \citet{2012A&A...538A..91D}, who predicted a similar gas-phase abundance for both O$_2$H and HOOH.  \citet{2012A&A...538A..91D} however have used dense conditions (density of $6\times 10^5$~cm$^{-3}$) with warm gas-phase temperature (21~K) based on observations. 
 These conditions are never reached in our simulations (dense is always cold in our model, see Fig.~\ref{points_dens_temp}). We tested a simple run with our chemical model, using the observed physical conditions (fixed with time) and were able to produce similar abundances for O$_2$H and HOOH, and similar to the observations and the predictions by \citet{2012A&A...538A..91D}. For these conditions at $3.6\times 10^4$~yr, we have an abundance of $2\times 10^{-10}$ for O$_2$H, $5\times 10^{-10}$ for HOOH, and $2\times 10^{-8}$ for O$_2$.

%\begin{figure}
%\includegraphics[width=0.475\linewidth]{density_clouds.pdf}
%\includegraphics[width=0.5\linewidth]{temperature_clouds.pdf}
%\caption{Total proton density (upper panel) and gas temperature (lower panel) as a function of time for case 1 (red dashed line) and case 2 (black solid line). \label{physics_clouds}}
%\end{figure}

\section{Predicted column densities}

\begin{figure*}
\includegraphics[width=0.3\linewidth]{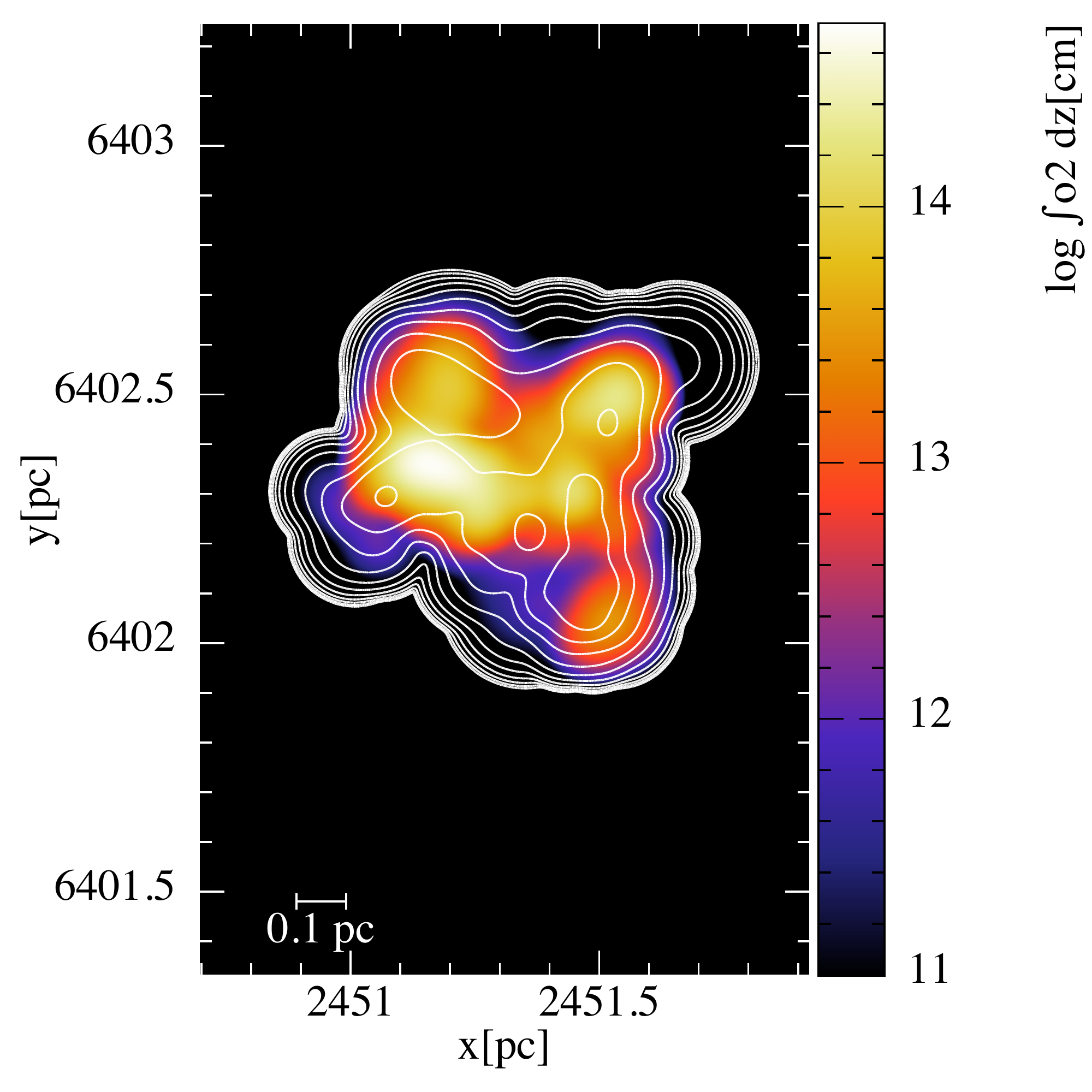}
\includegraphics[width=0.3\linewidth]{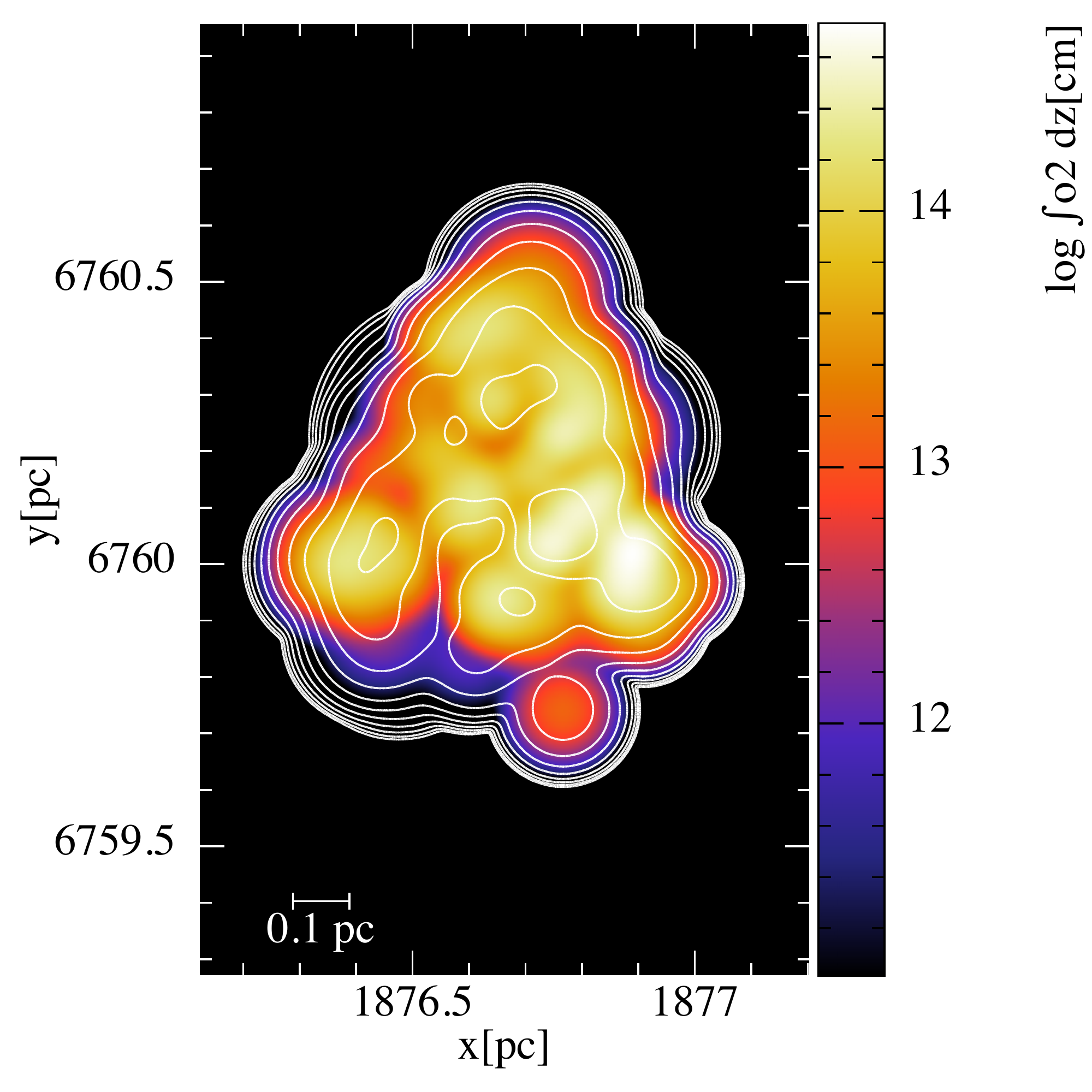}
\includegraphics[width=0.3\linewidth]{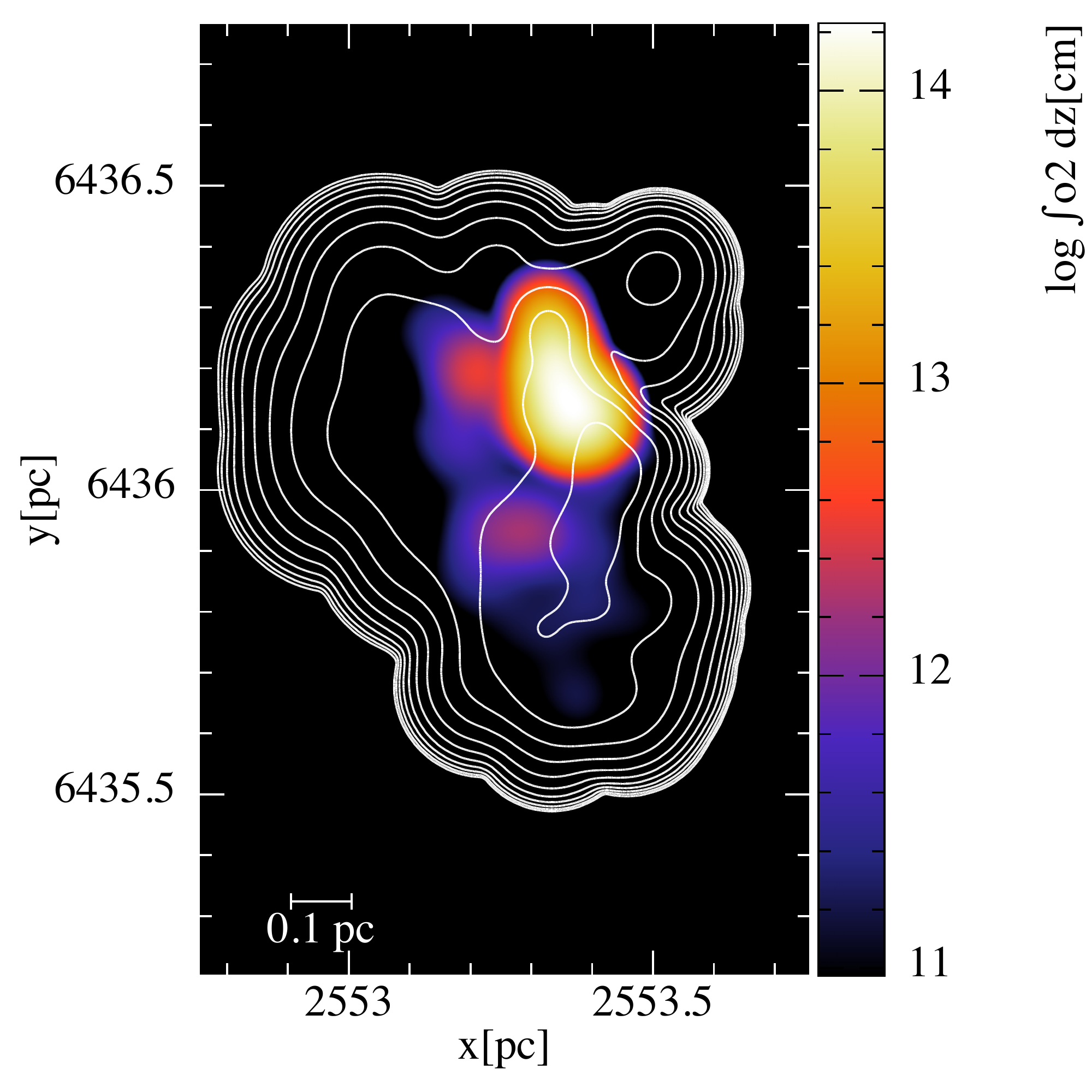}
\includegraphics[width=0.3\linewidth]{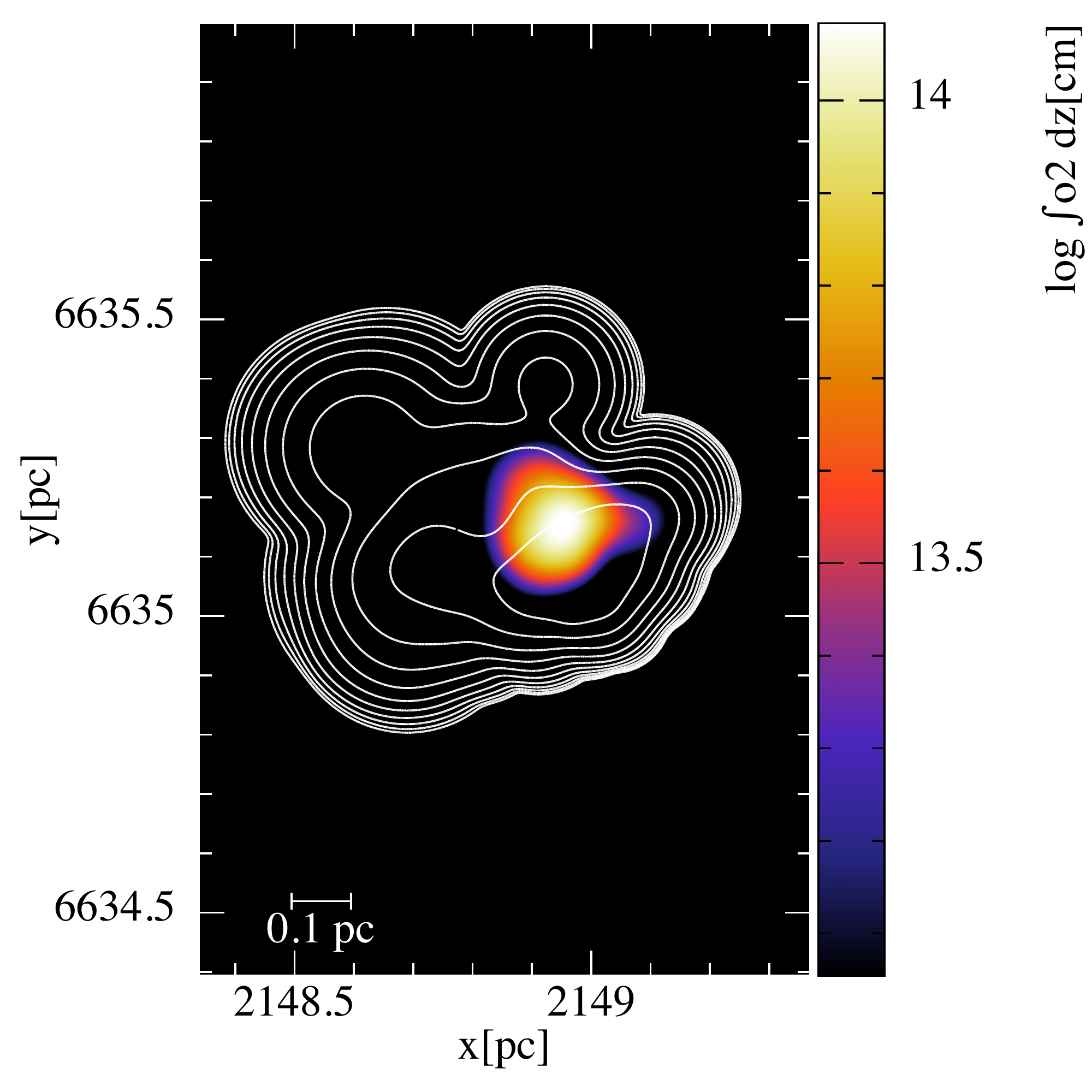}
\includegraphics[width=0.3\linewidth]{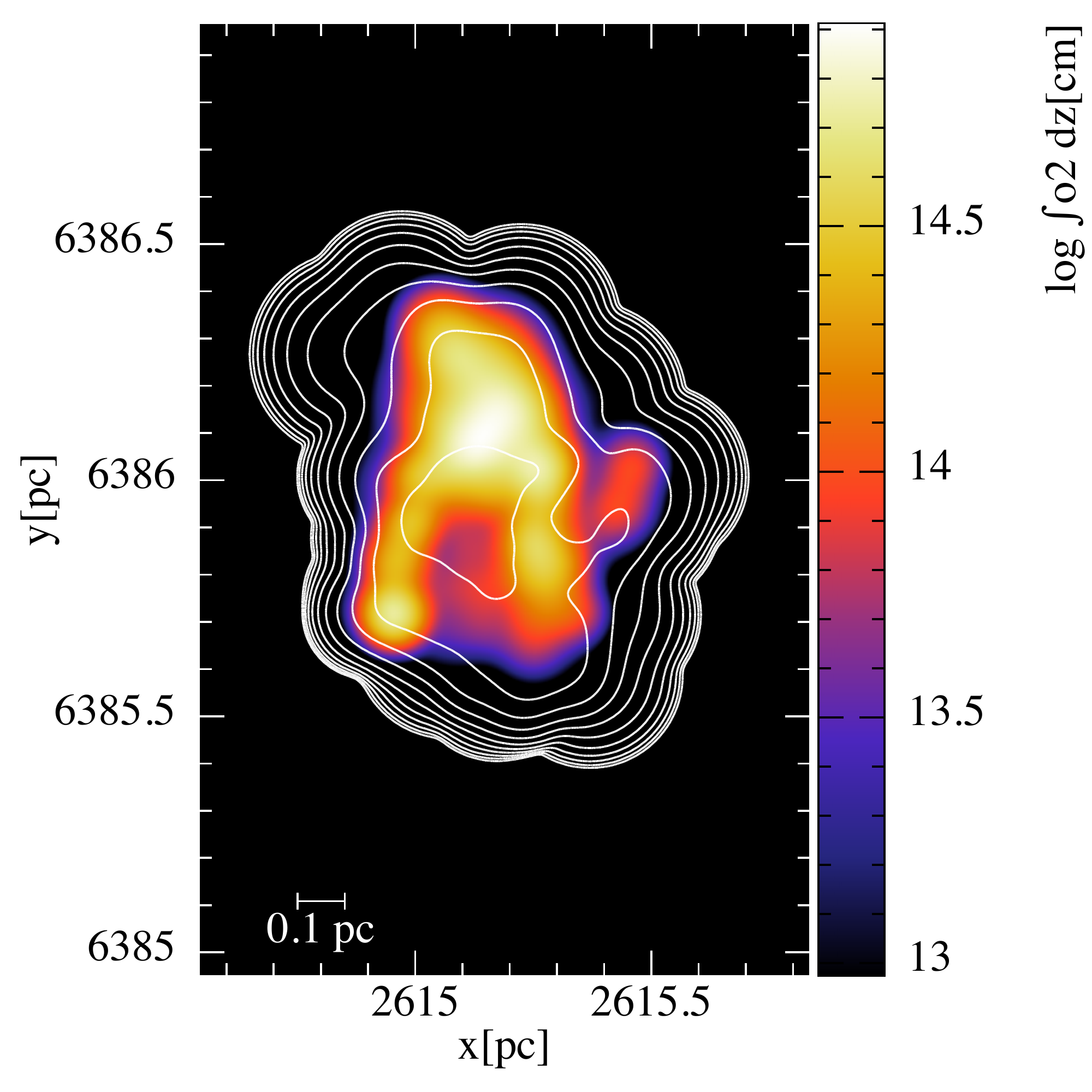}
\includegraphics[width=0.3\linewidth]{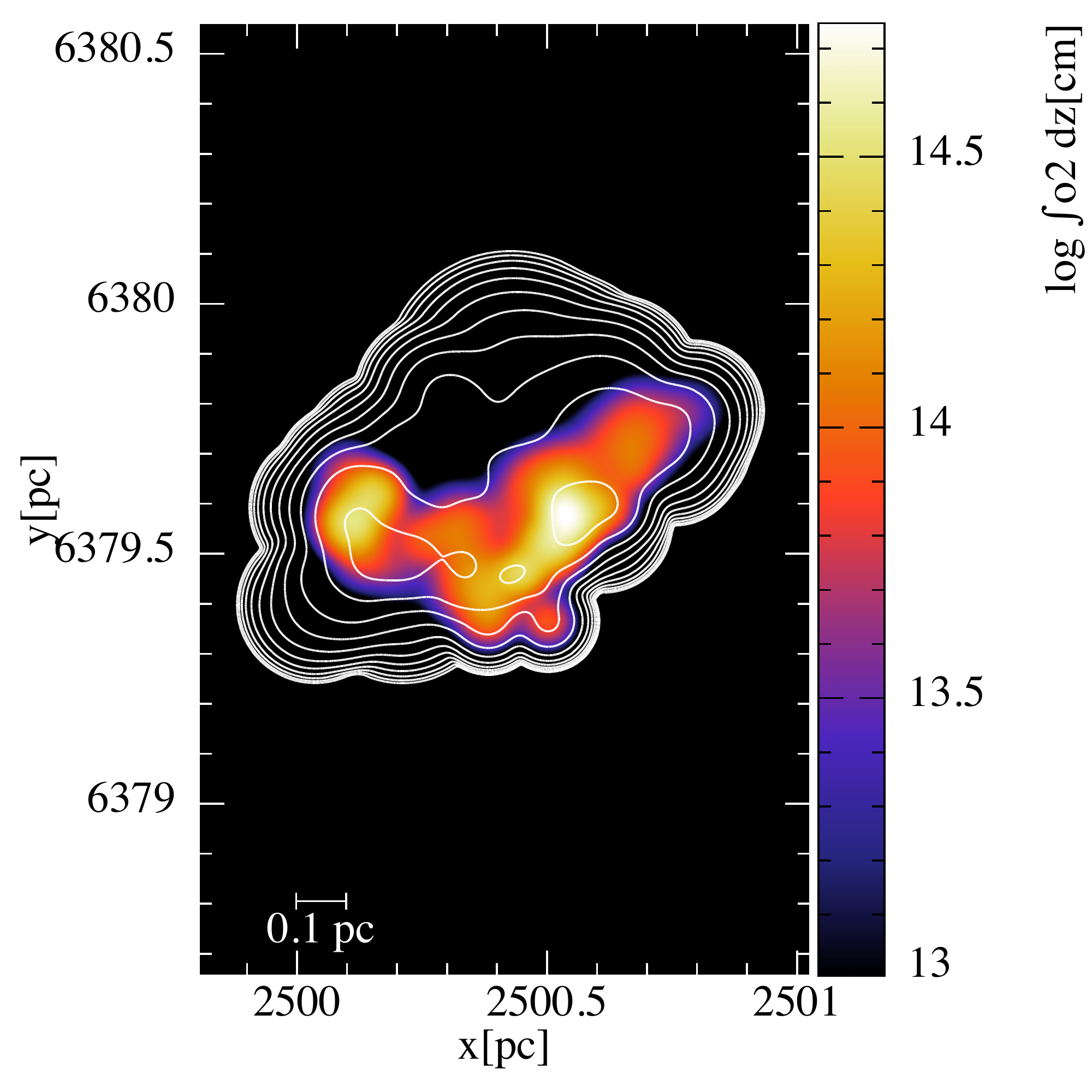}
\includegraphics[width=0.3\linewidth]{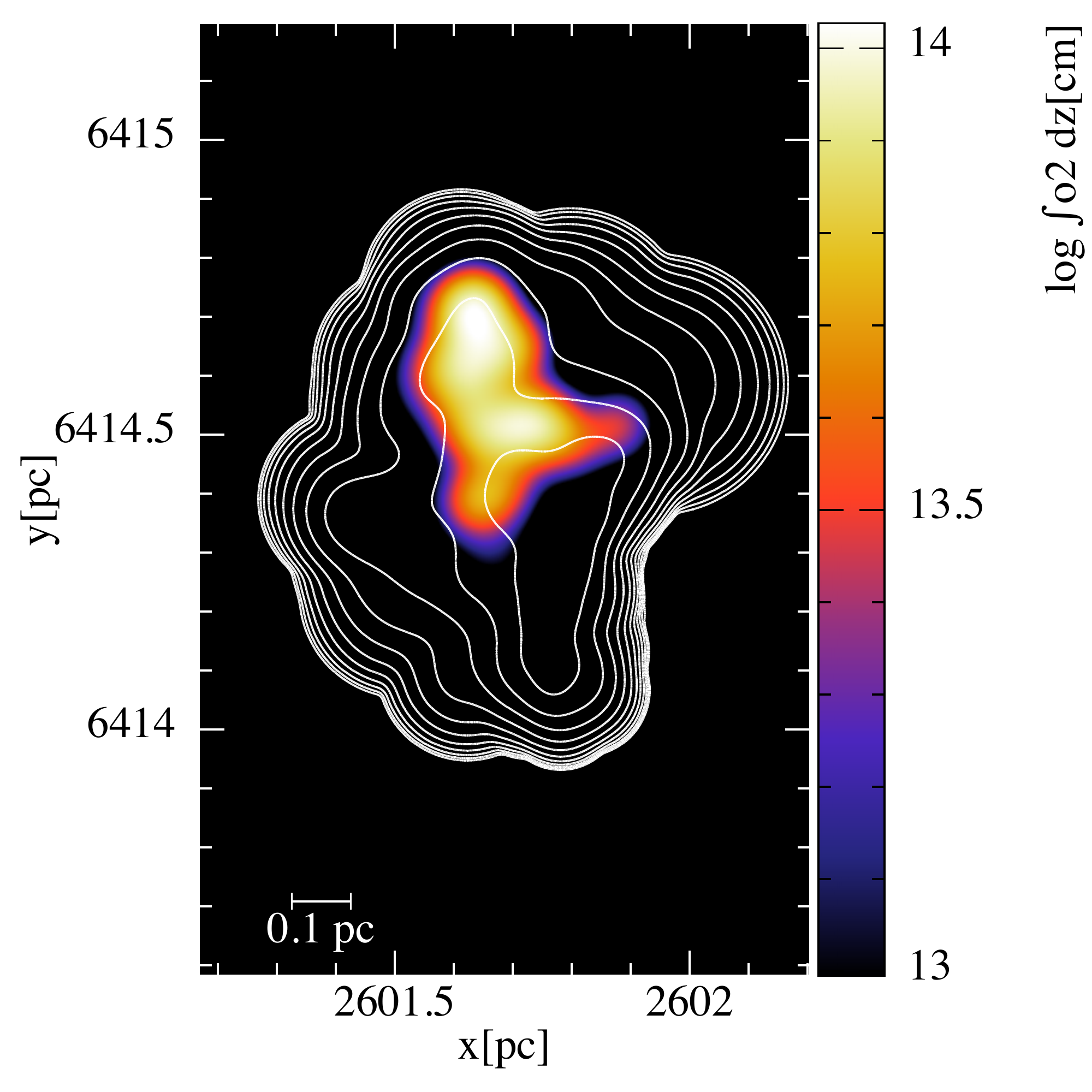}
\includegraphics[width=0.3\linewidth]{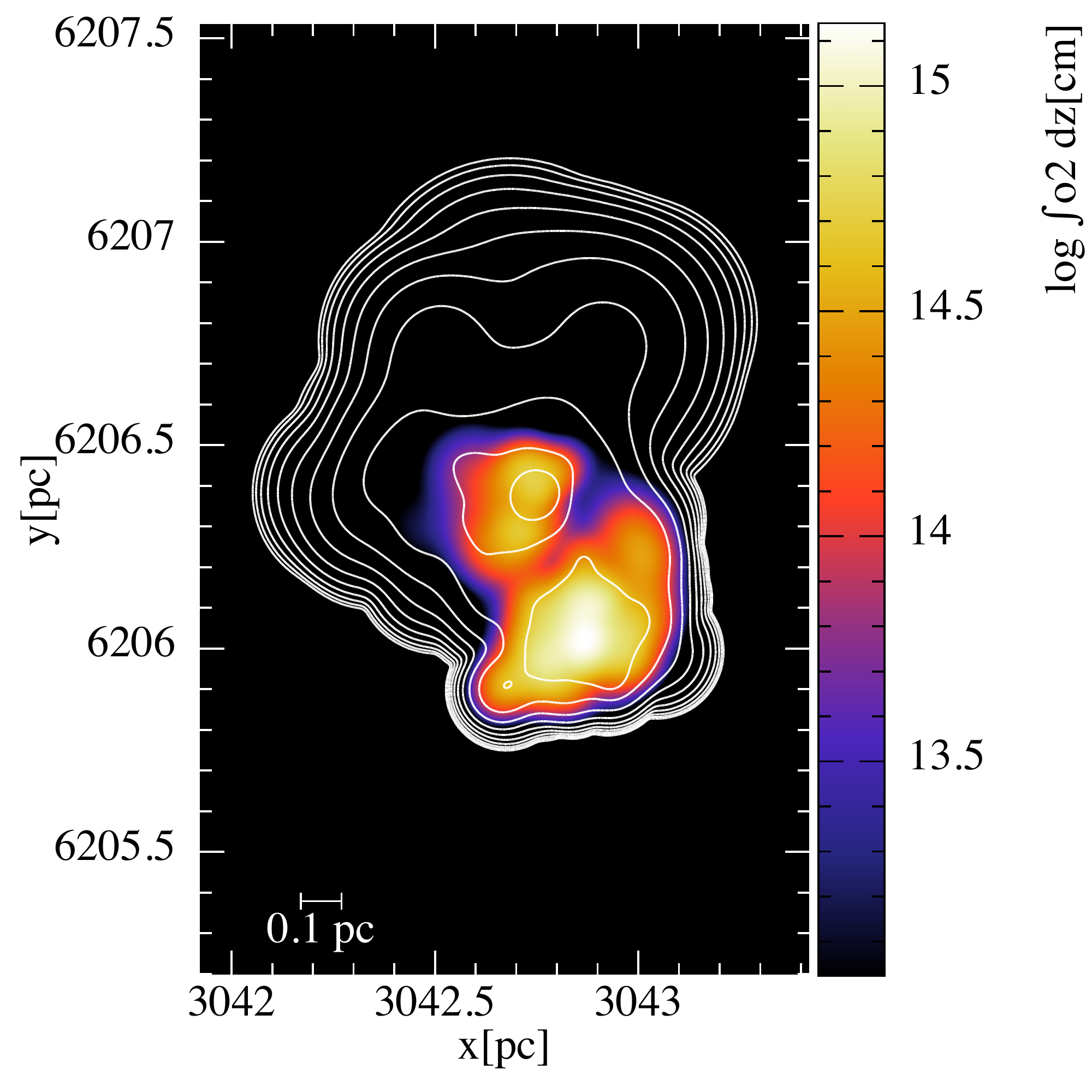}
\includegraphics[width=0.3\linewidth]{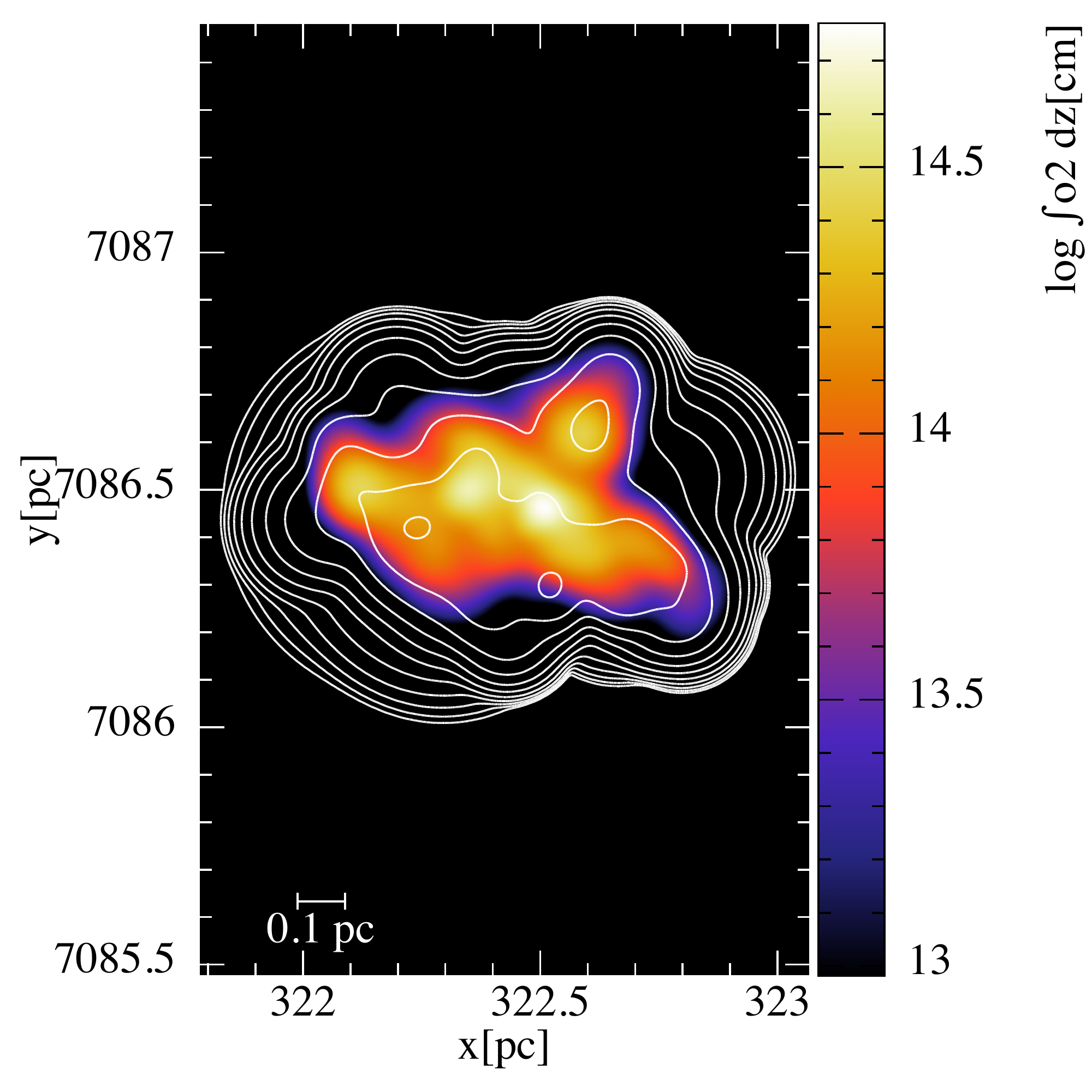}
\caption{O$_2$ predicted column density maps (in cm$^{-2}$) for nine clouds in color. The white contours are the total proton column density in log given every 0.5 dex and starting at 17.5. For all maps, the maximum peak contour is 22.5. \label{maps}}
\end{figure*}

To make a more reliable comparison between our predictions and the observations, we have created column density maps for our simulated clouds. For this, we have used the SPLASH visualisation tool (http://users.monash.edu.au/$\sim$dprice/splash/) \citep{2007PASA...24..159P}. Fig.~\ref{maps} shows the predicted column density maps of gas-phase O$_2$ in nine clouds. Color coding shows the log10 of the column density of O$_2$ while the white contours shows the contour levels of the log10 of the total proton density every 0.5 dex and starting at 17.5. Extension of the emission, as well as the distribution of O$_2$ within each cloud varies with the cloud. In all of them, the maximum column density is of a few $10^{14}$~cm$^{-3}$, in agreement with all the non detection limits listed in Table~\ref{table_obs}. This value is smaller than the O$_2$ column density observed in $\rho$ Oph A, but as we have shown earlier, this region has physical conditions not covered by our sample. As compared to the total proton density, it is interesting to notice that the O$_2$ does not peak at the same place as H in most cases. 

\section{Conclusion}

We report in this paper the predicted abundance of gas-phase O$_2$ computed by a full gas-grain model coupled with time dependent 
physical conditions from a SPH model for trajectories of material forming nine cold cores. Considering the 2913 trajectories 
covering a large parameter space, we found that the predicted O$_2$ abundance spreads over several orders of magnitudes for densities between $10^3$ and a few $10^5$ cm$^{-3}$ (resulting in temperatures between 10 and 75 K). Most of our predicted abundances are below $10^{-8}$. For the nine studied cold cores, the maximum O$_2$ predicted column density is of a few $10^{14}$~cm$^{-2}$. As compared to observations, these model results are in agreement with all upper limits of O$_2$ in cold cores. It is smaller than the column density observed in $\rho$ Oph A. In the same source, both O$_2$H and HOOH have been detected with a similar abundance. None of our models produce large abundances for the three molecules. We argue that the dense and warm conditions observed in $\rho$ Oph A are the reason for the large abundances of O$_2$H and HOOH, conditions which are never reached by our simulations. \\
As conclusion of this paper, we have shown that the rare detection of O$_2$ in the interstellar medium can be explained by the combined history of the interstellar material physical conditions and the time scale of the dynamical evolution. Indeed, in our simulations, the depletion of oxygen onto the grains (followed by hydrogenation) occurs before the gas-phase formation of O$_2$. When the peak density is reached, only a small fraction of oxygen is still available.

\section*{Acknowledgements}

VW's research is funded by an ERC Starting Grant (3DICE, grant agreement 336474). The authors acknowledge the CNRS program "Physique et Chimie du Milieu Interstellaire" (PCMI) co-funded by the Centre National d'Etudes Spatiales (CNES). IAB acknowledges funding from the European Research Council for the FP7 ERC advanced grant project ECOGAL. This work used the DiRAC Complexity system, operated by the University of Leicester IT Services, which forms part of the STFC DiRAC HPC Facility (www.dirac.ac.uk). This equipment is funded by BIS National E-Infrastructure capital grant ST/K000373/1 and STFC DiRAC Operations grant ST/K0003259/1. DiRAC is part of the National E-Infrastructure. MR's research was supported by an appointment to the NASA Postdoctoral Program at the NASA Ames Research Center, administered by Universities Space Research Association under contract with NASA.

%%%%%%%%%%%%%%%%%%%%%%%%%%%%%%%%%%%%%%%%%%%%%%%%%%



%\bibliographystyle{mnras}

\bibliographystyle{mnras}
\bibliography{bib}

\begin{thebibliography}{}
\makeatletter
\relax
\def\mn@urlcharsother{\let\do\@makeother \do\$\do\&\do\#\do\^\do\_\do\%\do\~}
\def\mn@doi{\begingroup\mn@urlcharsother \@ifnextchar [ {\mn@doi@}
  {\mn@doi@[]}}
\def\mn@doi@[#1]#2{\def\@tempa{#1}\ifx\@tempa\@empty \href
  {http://dx.doi.org/#2} {doi:#2}\else \href {http://dx.doi.org/#2} {#1}\fi
  \endgroup}
\def\mn@eprint#1#2{\mn@eprint@#1:#2::\@nil}
\def\mn@eprint@arXiv#1{\href {http://arxiv.org/abs/#1} {{\tt arXiv:#1}}}
\def\mn@eprint@dblp#1{\href {http://dblp.uni-trier.de/rec/bibtex/#1.xml}
  {dblp:#1}}
\def\mn@eprint@#1:#2:#3:#4\@nil{\def\@tempa {#1}\def\@tempb {#2}\def\@tempc
  {#3}\ifx \@tempc \@empty \let \@tempc \@tempb \let \@tempb \@tempa \fi \ifx
  \@tempb \@empty \def\@tempb {arXiv}\fi \@ifundefined
  {mn@eprint@\@tempb}{\@tempb:\@tempc}{\expandafter \expandafter \csname
  mn@eprint@\@tempb\endcsname \expandafter{\@tempc}}}

\bibitem[\protect\citeauthoryear{{Bergman}, {Parise}, {Liseau}, {Larsson},
  {Olofsson}, {Menten}  \& {G{\"u}sten}}{{Bergman}
  et~al.}{2011}]{2011A&A...531L...8B}
{Bergman} P.,  {Parise} B.,  {Liseau} R.,  {Larsson} B.,  {Olofsson} H.,
  {Menten} K.~M.,   {G{\"u}sten} R.,  2011, \mn@doi [\aap]
  {10.1051/0004-6361/201117170}, \href
  {http://adsabs.harvard.edu/abs/2011A%26A...531L...8B} {531, L8}

\bibitem[\protect\citeauthoryear{{Bonnell}, {Dobbs}  \& {Smith}}{{Bonnell}
  et~al.}{2013}]{2013MNRAS.430.1790B}
{Bonnell} I.~A.,  {Dobbs} C.~L.,   {Smith} R.~J.,  2013, \mn@doi [\mnras]
  {10.1093/mnras/stt004}, \href
  {http://adsabs.harvard.edu/abs/2013MNRAS.430.1790B} {430, 1790}

\bibitem[\protect\citeauthoryear{{Chen} et~al.,}{{Chen}
  et~al.}{2014}]{2014ApJ...793..111C}
{Chen} J.-H.,  et~al., 2014, \mn@doi [\apj] {10.1088/0004-637X/793/2/111},
  \href {http://adsabs.harvard.edu/abs/2014ApJ...793..111C} {793, 111}

\bibitem[\protect\citeauthoryear{{Du}, {Parise}  \& {Bergman}}{{Du}
  et~al.}{2012}]{2012A&A...538A..91D}
{Du} F.,  {Parise} B.,   {Bergman} P.,  2012, \mn@doi [\aap]
  {10.1051/0004-6361/201118013}, \href
  {http://adsabs.harvard.edu/abs/2012A%26A...538A..91D} {538, A91}

\bibitem[\protect\citeauthoryear{{Garrod}, {Wakelam}  \& {Herbst}}{{Garrod}
  et~al.}{2007}]{2007A&A...467.1103G}
{Garrod} R.~T.,  {Wakelam} V.,   {Herbst} E.,  2007, \mn@doi [\aap]
  {10.1051/0004-6361:20066704}, \href
  {http://adsabs.harvard.edu/abs/2007A%26A...467.1103G} {467, 1103}

\bibitem[\protect\citeauthoryear{{Goldsmith} et~al.,}{{Goldsmith}
  et~al.}{2000}]{2000ApJ...539L.123G}
{Goldsmith} P.~F.,  et~al., 2000, \mn@doi [\apjl] {10.1086/312854}, \href
  {http://adsabs.harvard.edu/abs/2000ApJ...539L.123G} {539, L123}

\bibitem[\protect\citeauthoryear{{Goldsmith} et~al.,}{{Goldsmith}
  et~al.}{2011}]{2011ApJ...737...96G}
{Goldsmith} P.~F.,  et~al., 2011, \mn@doi [\apj] {10.1088/0004-637X/737/2/96},
  \href {http://adsabs.harvard.edu/abs/2011ApJ...737...96G} {737, 96}

\bibitem[\protect\citeauthoryear{{Hasegawa} \& {Herbst}}{{Hasegawa} \&
  {Herbst}}{1993}]{1993MNRAS.261...83H}
{Hasegawa} T.~I.,  {Herbst} E.,  1993, \mn@doi [\mnras]
  {10.1093/mnras/261.1.83}, \href
  {http://adsabs.harvard.edu/abs/1993MNRAS.261...83H} {261, 83}

\bibitem[\protect\citeauthoryear{{Hincelin}, {Wakelam}, {Hersant},
  {Guilloteau}, {Loison}, {Honvault}  \& {Troe}}{{Hincelin}
  et~al.}{2011}]{2011A&A...530A..61H}
{Hincelin} U.,  {Wakelam} V.,  {Hersant} F.,  {Guilloteau} S.,  {Loison} J.~C.,
   {Honvault} P.,   {Troe} J.,  2011, \mn@doi [\aap]
  {10.1051/0004-6361/201016328}, \href
  {http://adsabs.harvard.edu/abs/2011A%26A...530A..61H} {530, A61}

\bibitem[\protect\citeauthoryear{{Hollenbach}, {Takahashi}  \&
  {Tielens}}{{Hollenbach} et~al.}{1991}]{1991ApJ...377..192H}
{Hollenbach} D.~J.,  {Takahashi} T.,   {Tielens} A.~G.~G.~M.,  1991, \mn@doi
  [\apj] {10.1086/170347}, \href
  {http://adsabs.harvard.edu/abs/1991ApJ...377..192H} {377, 192}

\bibitem[\protect\citeauthoryear{{Larsson} et~al.,}{{Larsson}
  et~al.}{2007}]{2007...466..999L}
{Larsson} B.,  et~al., 2007, \mn@doi [\aap] {10.1051/0004-6361:20065500}, \href
  {http://adsabs.harvard.edu/abs/2007A%26A...466..999L} {466, 999}

\bibitem[\protect\citeauthoryear{{Liseau} et~al.,}{{Liseau}
  et~al.}{2012}]{2012...541A..73L}
{Liseau} R.,  et~al., 2012, \mn@doi [\aap] {10.1051/0004-6361/201118575}, \href
  {http://adsabs.harvard.edu/abs/2012A%26A...541A..73L} {541, A73}

\bibitem[\protect\citeauthoryear{{Loison} et~al.,}{{Loison}
  et~al.}{2017}]{2017MNRAS.470.4075L}
{Loison} J.-C.,  et~al., 2017, \mn@doi [\mnras] {10.1093/mnras/stx1265}, \href
  {http://adsabs.harvard.edu/abs/2017MNRAS.470.4075L} {470, 4075}

\bibitem[\protect\citeauthoryear{{Marechal}, {Viala}  \& {Benayoun}}{{Marechal}
  et~al.}{1997}]{1997A&A...324..221M}
{Marechal} P.,  {Viala} Y.~P.,   {Benayoun} J.~J.,  1997, \aap, \href
  {http://adsabs.harvard.edu/abs/1997A%26A...324..221M} {324, 221}

\bibitem[\protect\citeauthoryear{{Melnick} et~al.,}{{Melnick}
  et~al.}{2012}]{2012ApJ...752...26M}
{Melnick} G.~J.,  et~al., 2012, \mn@doi [\apj] {10.1088/0004-637X/752/1/26},
  \href {http://adsabs.harvard.edu/abs/2012ApJ...752...26M} {752, 26}

\bibitem[\protect\citeauthoryear{{Pagani} et~al.,}{{Pagani}
  et~al.}{2003}]{Pagani2003}
{Pagani} L.,  et~al., 2003, \mn@doi [\aap] {10.1051/0004-6361:20030344}, \href
  {http://adsabs.harvard.edu/abs/2003A%26A...402L..77P} {402, L77}

\bibitem[\protect\citeauthoryear{{Parise}, {Bergman}  \& {Du}}{{Parise}
  et~al.}{2012}]{2012A&A...541L..11P}
{Parise} B.,  {Bergman} P.,   {Du} F.,  2012, \mn@doi [\aap]
  {10.1051/0004-6361/201219379}, \href
  {http://adsabs.harvard.edu/abs/2012A%26A...541L..11P} {541, L11}

\bibitem[\protect\citeauthoryear{{Price}}{{Price}}{2007}]{2007PASA...24..159P}
{Price} D.~J.,  2007, \mn@doi [\pasa] {10.1071/AS07022}, \href
  {http://adsabs.harvard.edu/abs/2007PASA...24..159P} {24, 159}

\bibitem[\protect\citeauthoryear{{Quan}, {Herbst}, {Millar}, {Hassel}, {Lin},
  {Guo}, {Honvault}  \& {Xie}}{{Quan} et~al.}{2008}]{2008ApJ...681.1318Q}
{Quan} D.,  {Herbst} E.,  {Millar} T.~J.,  {Hassel} G.~E.,  {Lin} S.~Y.,  {Guo}
  H.,  {Honvault} P.,   {Xie} D.,  2008, \mn@doi [\apj] {10.1086/588007}, \href
  {http://adsabs.harvard.edu/abs/2008ApJ...681.1318Q} {681, 1318}

\bibitem[\protect\citeauthoryear{{Ruaud}, {Wakelam}  \& {Hersant}}{{Ruaud}
  et~al.}{2016}]{2016MNRAS.459.3756R}
{Ruaud} M.,  {Wakelam} V.,   {Hersant} F.,  2016, \mn@doi [\mnras]
  {10.1093/mnras/stw887}, \href
  {http://adsabs.harvard.edu/abs/2016MNRAS.459.3756R} {459, 3756}

\bibitem[\protect\citeauthoryear{{Ruaud}, {Wakelam}, {Gratier}  \&
  {Bonnell}}{{Ruaud} et~al.}{2018}]{2018A&A...611A..96R}
{Ruaud} M.,  {Wakelam} V.,  {Gratier} P.,   {Bonnell} I.~A.,  2018, \mn@doi
  [\aap] {10.1051/0004-6361/201731693}, \href
  {http://adsabs.harvard.edu/abs/2018A%26A...611A..96R} {611, A96}

\bibitem[\protect\citeauthoryear{{Sandqvist} et~al.,}{{Sandqvist}
  et~al.}{2015}]{Sandqvist2015}
{Sandqvist} A.,  et~al., 2015, \mn@doi [\aap] {10.1051/0004-6361/201526280},
  \href {http://adsabs.harvard.edu/abs/2015A%26A...584A.118S} {584, A118}

\bibitem[\protect\citeauthoryear{{Taquet} et~al.,}{{Taquet}
  et~al.}{2018}]{2018...618A..11T}
{Taquet} V.,  et~al., 2018, \mn@doi [\aap] {10.1051/0004-6361/201833175}, \href
  {http://adsabs.harvard.edu/abs/2018A%26A...618A..11T} {618, A11}

\bibitem[\protect\citeauthoryear{{Viti}, {Roueff}, {Hartquist}, {Pineau des
  For{\^e}ts}  \& {Williams}}{{Viti} et~al.}{2001}]{2001A&A...370..557V}
{Viti} S.,  {Roueff} E.,  {Hartquist} T.~W.,  {Pineau des For{\^e}ts} G.,
  {Williams} D.~A.,  2001, \mn@doi [\aap] {10.1051/0004-6361:20010246}, \href
  {http://adsabs.harvard.edu/abs/2001A%26A...370..557V} {370, 557}

\bibitem[\protect\citeauthoryear{{Wakelam} et~al.,}{{Wakelam}
  et~al.}{2015}]{2015ApJS..217...20W}
{Wakelam} V.,  et~al., 2015, \mn@doi [\apjs] {10.1088/0067-0049/217/2/20},
  \href {http://adsabs.harvard.edu/abs/2015ApJS..217...20W} {217, 20}

\bibitem[\protect\citeauthoryear{{Wakelam}, {Loison}, {Mereau}  \&
  {Ruaud}}{{Wakelam} et~al.}{2017}]{2017MolAs...6...22W}
{Wakelam} V.,  {Loison} J.-C.,  {Mereau} R.,   {Ruaud} M.,  2017, \mn@doi
  [Molecular Astrophysics] {10.1016/j.molap.2017.01.002}, \href
  {http://adsabs.harvard.edu/abs/2017MolAs...6...22W} {6, 22}

\bibitem[\protect\citeauthoryear{{Wirstr{\"o}m}, {Charnley}, {Cordiner}  \&
  {Ceccarelli}}{{Wirstr{\"o}m} et~al.}{2016}]{2016ApJ...830..102W}
{Wirstr{\"o}m} E.~S.,  {Charnley} S.~B.,  {Cordiner} M.~A.,   {Ceccarelli} C.,
  2016, \mn@doi [\apj] {10.3847/0004-637X/830/2/102}, \href
  {http://adsabs.harvard.edu/abs/2016ApJ...830..102W} {830, 102}

\bibitem[\protect\citeauthoryear{{Y{\i}ld{\i}z} et~al.,}{{Y{\i}ld{\i}z}
  et~al.}{2013}]{2013...558A..58Y}
{Y{\i}ld{\i}z} U.~A.,  et~al., 2013, \mn@doi [\aap]
  {10.1051/0004-6361/201321944}, \href
  {http://adsabs.harvard.edu/abs/2013A%26A...558A..58Y} {558, A58}

\makeatother
\end{thebibliography}

%%%%%%%%%%%%%%%%%%%%%%%%%%%%%%%%%%%%%%%%%%%%%%%%%%

%%%%%%%%%%%%%%%%% APPENDICES %%%%%%%%%%%%%%%%%%%%%

%\appendix



%%%%%%%%%%%%%%%%%%%%%%%%%%%%%%%%%%%%%%%%%%%%%%%%%%

% Don't change these lines
\bsp	% typesetting comment
\label{lastpage}
\end{document}